\documentclass[preprintnumbers,amsmath,amssymb]{revtex4}

\usepackage{graphicx}
\usepackage{dcolumn}
\usepackage{bm}

\begin{document}

\title{Loss of Cell-Substrate Adhesion  Leads to Periodic Shape Oscillations in
Fibroblasts}

\author{Pramod A. Pullarkat}
\altaffiliation[]{
email: pramod.pullarkat@uni-bayreuth.de; mail2pramod@gmail.com
}
\affiliation{Physikalisches Institut, University of
Bayreuth, D95440-Bayreuth, Germany.}
\date{\today}

\begin{abstract}
We report the phenomenon of periodic shape oscillations occurring in
3T3 fibroblasts merely as a consequence of a loss of cell-substrate adhesion.
The oscillatory behavior can last many hours at a constant frequency, and can be
switched off and on using chemical agents. This robustness allows for the
extraction of quantitative data using single cells. We demonstrate that the 
frequency of oscillations increases with increasing actomyosin contractility.
Both the Myosin Light Chain Kinase as well as the Rho Kinase pathways are shown
to operate during this process.
Further, we reveal significant similarities between the oscillatory dynamics and
the commonly observed phenomenon of blebbing. We show that both these processes
ceases when extracellular calcium is depleted or when stretch activated calcium
channels are blocked. This, along with the fact that these dynamical
processes require actomyosin contactility points towards strong similarities
in the respective mechanisms. Finally, we speculate on a possible mechanism for
the observed dynamical phenomena.
\end{abstract}
\maketitle

\section*{INTRODUCTION}

\vspace{-0.4cm}
The cell cytoskeleton is a biopolymer gel capable of generating active
contractile forces and rapid reorganization ~\cite{alberts, bray}. These
properties are critical for the ability of these filament networks, especially
the actomyosin gel, in controlling cell shape and vital functions like
locomotion, cytokinesis, mechanotransduction, etc ~\cite{alberts, bray}. The
active remodeling of the cytoskeleton during such dynamical processes involves
complex signal transduction pathways which include mechano-chemical coupling.
Of the myriad of functions the cytoskeleton performs in cells, oscillatory
dynamics observed in certain cell types are of particular interest to biologists
and physicists alike. Cells like cardiomyocytes and asynchronous
insect muscle cells have evolved to generate rhythmic contractile
forces ~\cite{hescheler1999, pringle, jacques1997, frank2001}. 
It is also known that certain non-oscillatory cells can be induced to exhibit
oscillatory behavior by biochemical intervention. Examples are shape
oscillations occurring in lymphoblasts, spreading fibroblasts and in cell
fragments, as a result of microtubule depolymerization using drugs
~\cite{bornens1989, pletjushkina2001, paluch2005}. Shape oscillations
are also observed in neutrophils and amoeboid cells, usually when
exposed to chemoattractants ~\cite{deranleau1995, kobatake1985}
and in the lamellipodea of spreading cells ~\cite{sheetz2004, danuser2006}.
Recently, spontaneous oscillations of isolated myofibrils
 lacking regulatory proteins have been observed
and theoretically studied ~\cite{ishiwata1998, jacques1997}.  Although
oscillatory dynamics in simpler systems like beating flagella are well explained
by physical models ~\cite{jacques1999, frank2001}, gaining a theoretical
understanding of the dynamics of actomyosin gel is far more complicated and
several aspects are still a challenge, despite considerable progress in the
field ~\cite{oster1984, jacques1997, yhat2004, jacques2006, frank_oscil_2005}.

In this article, we report the observation of  oscillatory shape dynamics
in freely suspended Swiss-3T3 fibroblasts, without any treatment using drugs. 
Remarkably, this phenomenon arises merely as a result of loss of cell-substrate
adhesion. By maintaining the cells in suspension, we are able to show that the
frequency of oscillation can remain constant for several hours allowing
quantitative measurements. We show that the frequency is decreased when
myosin motor activity is reduced using drugs.
Both the Myosin Light Chain Kinase (MLCK) and the Rho kinase pathways play a
role in the myosin activation process.
Further, we show that a small
minimum level of extracellular calcium is essential for oscillations and that
this calcium enters the cell via mechanosensitive calcium channels. Later, we
compare the oscillatory dynamics with the commonly observed blebbing dynamics
and reveal some striking similarities between these two seemingly different
phenomena. As detachment from the surface is enough to trigger oscillations, we
conclude that this dynamics constitute a fundamental response of the cortical
actomyosin gel in these cells and may occur under physiological conditions.
Finally, we speculate on possible mechanisms for the observed phenomena.

\section*{MATERIALS AND METHODS}

Swiss-3T3 fibroblasts were obtained from DSMZ Germany ~\cite{DSMZ}  and cultured
following a standard protocol. The adherent cells, grown to 50\% to 80\%
confluence in plastic cell culture flasks were washed with HBSS devoid of
$\rm{Ca}^{++}$ and $\rm{Mg}^{++}$ and then treated with a minimal amount of
0.25\% Trypsin solution for less than 5 minutes in order to
detach them from the cell culture flask. In other cases, mitotic cells were
detached by gentle shake-off, without the use of Trypsin. The cells were
immediately transferred to a petri dish with a glass coverslip bottom and Iscove
(IMDM) cell culture medium with 25 mM HEPES buffer and 10\% heat inactivated
Fetal Bovine Serum. All cell culture reagents were obtained from
Gibco-Invitrogen. Observations were made using a Zeiss Axiovert-135 microscope
configured for phase-contrast observation. The temperature was controlled using
a home-built temperature controller within $\pm{0.1}$ $^\circ$C.

Hydrophobic surfaces which inhibit cell-substrate adhesion were obtained by
coating clean glass coverslips with dimethyldichlorosilane by evaporation and
subsequent rinsing using ethanol. In other cases a clean glass coverslip was
used to study the effect of cell spreading on oscillatory behavior. Since the
cells were held in suspension in most experiments, we maintained a small upward
temperature gradient of about 0.1 $^\circ$C/mm in the petridish using
differential heating in order to avoid convective instabilities. This gradient
did not have any observable effect on the oscillations.

The images of the shape changes occurring in the fibroblasts were recorded using
a Spot-RT CCD camera (Diagnostic instruments Inc). The 2-dimensional projections
were analyzed using a home-developed image analysis program. Typically, the
bright edges in the phase contrast images were detected by applying an intensity
thresholding. Either the enclosed area or the position of the centroid was used
to measure the frequency of oscillations, depending on which of these quantities
exhibited clearly detectable oscillations. In cases where oscillations are
detected in both these quantities, they have the same frequency.
Fluorescence recordings were performed using a high gain EM-CCD camera from
Hamamatsu and a 60X, 1.4 NA objective.

The stock solutions of the chemicals were prepared as follows: 0.2 mg/ml
Latrunculin-A, 0.37 mM (-)-blebbistatin, and 50 mM ML-7 were prepared 
in dimethylsulphoxide (DMSO); 100 mM Ethylenediaminetetraacetic acid (EDTA), 
1 mM Gadolinium solution using Gadolinium Chloride Hydrate, and 10 mM Y-27632
were prepared in deionized water; 1 mM Lisophosphatidic acid was prepared in PBS
buffer (all chemicals are from Sigma). 1 mM Fluo-4 AM in DMSO (Invitrogen,
Molecular Probes) was used directly for imaging intracellular free calcium.
These stock solutions were added to the culture medium to obtain the desired
final concentrations. For calcium imaging, 1 $\mu$l Fluo-4 was added to adherent
cells in a culture flask containing 2 ml IMDM medium and 1\% serum, and left at
room temperature for 10 to 20 minutes. The excess dye was then washed off and
the cells were incubated for an hour before use.

Specifically coated beads were prepared by rinsing 4.5 $\mu$m
Dynabeads M-450 Epoxy (Dynal) in PBS three times and suspending them in a
solution of Fibronectin (0.1 \%, Sigma, further diluted by 100 times in PBS) for
1 hour. The suspension was maintained in a vertical rotator to prevent clumping
of the beads and later added to the cells in culture.

\section*{RESULTS}

\begin{figure}[h]
\includegraphics[width=15.0cm]{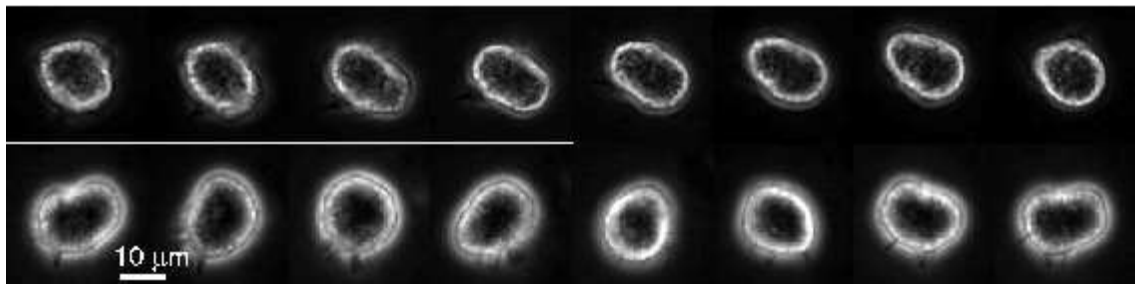}
\vspace{-0.3cm}
\caption{\label{fig:seq1}
Spontaneously occurring shape oscillations in non-adherent 3T3 fibroblasts.
Image sequences corresponds to one oscillation period of a small cell
at 3 s/frame (above), and of a larger one at 5 s/frame (below). 
The smaller cell exhibits an elliptic-like mode, whereas the
bigger one shows a more complex shape dynamics, although also
periodic. Also see supplementary movies Movie-1 and Movie-2.
}
\vspace{-0.0cm}
\end{figure}
\begin{figure}[h]
\includegraphics[width=15.0cm]{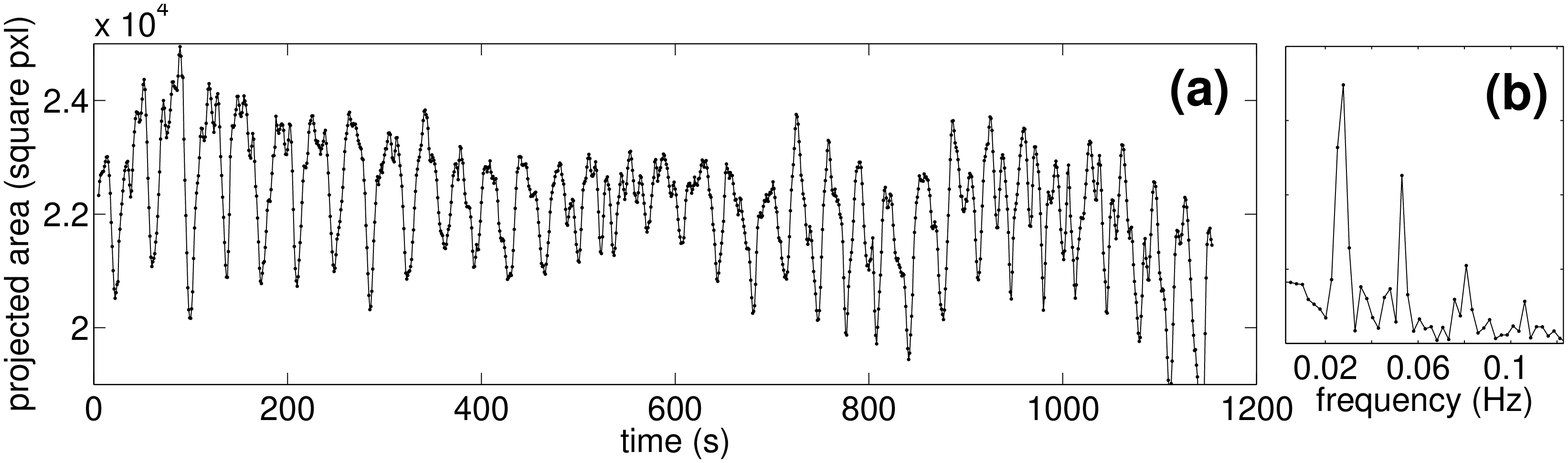}
\vspace{-0.3cm}
\caption{\label{fig:c11}
(a) Oscillations in the projected area of the 3T3 fibroblast shown in 
the lower sequence of Fig.\ \ref{fig:seq1}. The cell was maintained
at 35 ${^\circ}$C. The oscillations lasted many hours without any
significant change in frequency. 
(b) Power spectrum of the data in (a) showing peaks at the main frequency
of 0.028 Hz and its higher harmonics.
}
\vspace{-0.0cm}
\end{figure}

\begin{figure}[h]
\includegraphics[width=10.0cm]{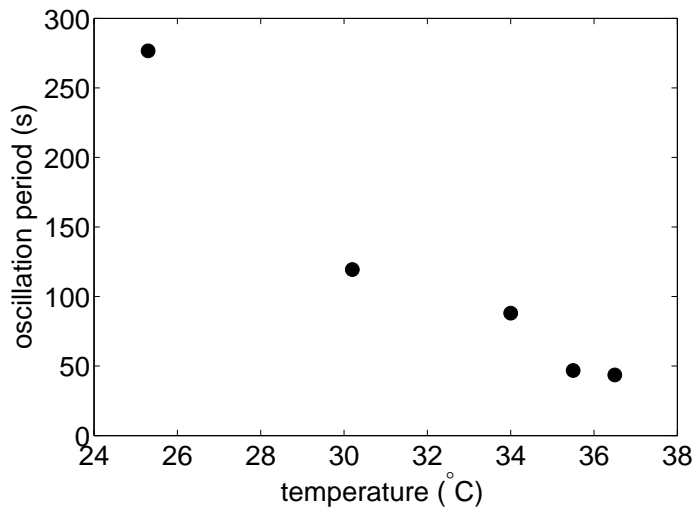}
\vspace{0.0cm}
\caption{\label{fig:freq-temp}
Variation in oscillation frequency as a function of temperature
obtained using a single cell.
}
\vspace{0.0cm}
\end{figure}

{\bf Shape oscillations in cells:} When the cells are maintained in suspension
using hydrophobically treated glass substrates, about 40\% to 60\% of them show
continuous shape alterations. Most of these cells exhibit an oscillatory
behavior as shown in Fig.\ \ref{fig:seq1}. The periodic
behavior can be much better perceived from the supplementary 
movies Movie-1 and Movie-2 (available at \cite{movies}). Image
analysis of the shape reveals a highly periodic dynamics as shown in
Fig.\ \ref{fig:c11}. On hydrophobic surfaces constant frequency oscillations can
then be observed for several hours. When oscillating cells are left at room
temperature overnight and heated to the normal working temperature (35
${^\circ}$C)  the next day, the oscillations resume with almost the same
frequency as before. This shows that the oscillatory behavior is a very robust
feature of these cells. The period of oscillations, however, depends on
temperature as shown in Fig.\ \ref{fig:freq-temp}. It may be noted that
all metabolic processes, including motor protein activity and filament dynamics,
are strongly temperature dependent.

Different dynamical behaviors can be observed among active cells, in general
depending on the cell size. (i) Small cells tend to show very periodic
behavior. The oscillations in them often show an elliptic-like mode with the
cell tending to oscillate along a preferred direction as shown in 
Fig.\ \ref{fig:seq1} (upper sequence). (ii) Larger cells often show very
periodic dynamics but the oscillation modes are more complex as is the case in
Fig.\ \ref{fig:seq1} (lower sequence). No clearly defined symmetries could
be observed in them. Image analysis shows that in some cases the shape change
appears as a propagating cortical wave. (iii) Among larger cells, a few tend to
show highly dynamical behavior but without any well defined periodicity in any
of the analyzed quantities (see supplementary movie Movie-3 \cite{movies}). In
a population of oscillating cells, the period ranges from about 25 s to 45 s
without any clear dependence on cell size. The cell size varies
only by about a factor of two making it difficult to quantify these
tendencies in behavior with size. Determination of volume of the
oscillating cells is also difficult due to the fact that the observed shapes, in
general, do not have any symmetry properties that can be exploited to estimate
the volume. With this in mind, no clear indication of any volume change could be
observed in small cells with less complex shapes.

{\bf Effect of cell-substrate adhesion:}
Since the oscillations are observed in cells which are detached from the
substrate, we expored the role of adhesion on the oscillatory process and find
the following. (i) Cells which are not attached to the hydrophobic
substrates (free floating) and those which have point-like contacts behave
identically as far as the oscillatory dynamics is concerned. The absence or
existence of adhesion is easily verified by observing the diffusive motion of
the cells and/or by applying a small flow disturbance to the culture medium.
(ii) When cells are seeded on clean, untreated glass substrates (a non-specific
substrate), they attach and eventually spread. The oscillations are observed
before the cells attach and after attachment. As cell spreading proceeds the
oscillations become progressively weaker and eventually dies out. In cases where
the cell subsequently detaches from the surface on its own, the oscillations
resume with almost the same frequency as before. (iii) When an active cell is
observed to drift and make contact with another cell, forming an adhesion patch
of about 5 $\mu$m, the dynamics continues without any observable change. (iv)
Non-specific adhesions are different from specific adhesions to adhesion
promoting proteins as far as signaling and cytoskeletal organization are
concerned. For this reason we used 4.5 micron epoxy beads coated with
Fibronectin and let them attach to the cells. The adhesion is verified by
pulling on beads stuck to partly spread cells using a home built magnetic
tweezer. When these cells are observed in suspension one to three beads are
seen firmly sticking to most of the cells. These cells exhibit oscillations
with almost the same probability as control cells. 

It may be assumed that detachment of the cell-substrate contacts triggers
signal transduction pathways which lead to oscillations. However,
the above mentioned experiments clearly show that having no cell-substrate
contact or having small adhesive contacts to specific or non-specific substrates
do not affect the observed dynamics. Importantly, whether oscillations are
observed or not does depend on the extent of adhesion suggesting mechanical
constraints may impede oscillations (see point (ii) mentioned above). 

For a given cell, the period of oscillation can remain constant for hours. In
what follows, we take advantage of this fact to investigate the
response of individual cells to biochemical perturbations using drugs and other
chemical agents.

{\bf Effect of actomyosin contractility on oscillations:}
\begin{figure}[h]
\includegraphics[width=10.0cm]{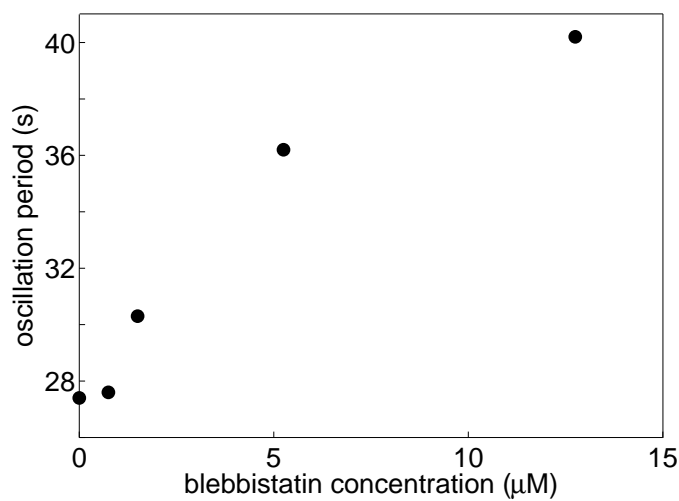}
\vspace{-0.3cm}
\caption{\label{fig:period_blebbi}
Oscillation period of a given cell as a function of concentration of 
blebbistatin at 35.5 $^\circ$C. The oscillations become weak, erratic
and intermittent towards the highest concentration.
}
\vspace{-0.0cm}
\end{figure}
%
We observe that depolymerization of actin filaments using Latrunculin-A or
deactivating myosin using blebbistatin kills the above described dynamical
behavior. In order to quantify how the activity of cortical actomyosin gel
influences the dynamics we exposed the cells to increasing concentration of
blebbistatin, a highly specific inhibitor of myosin II
~\cite{mitchison2003-blebbi, korn2005}.
We observe a clear dependence of the oscillation period on the
concentration of the drug as shown in Fig.\ \ref{fig:period_blebbi}. This
implies a direct relation between the activity of myosin motors and
the frequency of oscillations. At drug concentrations of about 5 $\mu$M the
oscillations are observed only intermittently. At still higher
concentrations the shape dynamics ceases completely.

\begin{figure}[h]
\includegraphics[width=10.0cm]{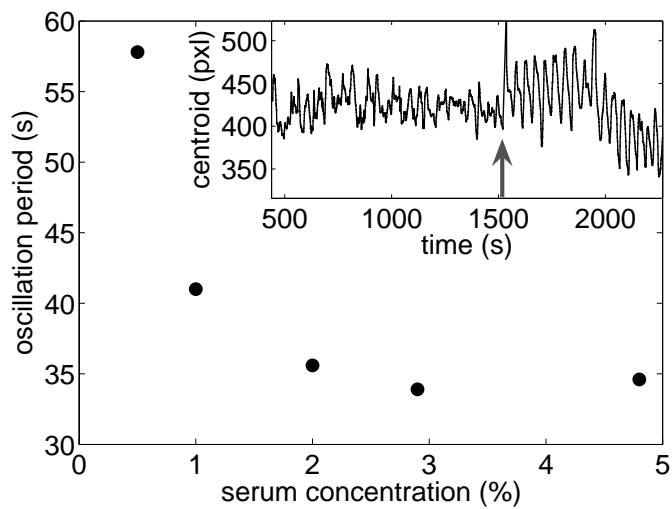}
\vspace{-0.3cm}
\caption{\label{fig:period_serum}
Oscillation period of a given cell as a function of concentration of 
serum at 35.5 $^\circ$C. The inset shows a remarkable transition from
erratic oscillations to regular oscillations as the serum concentration is
increased from 0.5\% to 1\% (arrow).
}
\vspace{-0.0cm}
\end{figure}

In order to perform a complementary experiment to verify the
effect of contractility on the period, we exposed normal cells to
increasing amounts of serum. Serum is known to increase
contractility in fibroblasts in a linear fashion with concentration 
\cite{obara2000}, due to increased myosin phosphorylation
~\cite{obara2003,serum_taylor1992}. Indeed, we observe a decrease in
oscillation period with increasing serum concentration as shown in Fig.\
\ref{fig:period_serum}.
%

These experiments show that the frequency of oscillations increases with
increasing actomyosin contractility. Remarkably, there is a clear transition
from erratic shape dynamics to periodic oscillations as the contractility is
increased as shown in Fig.\ \ref{fig:period_serum}(inset). Erratic oscillations
are also observed at high blebbistatin concentrations (about 5 $\mu$M),
suggesting a dynamic transition from erratic to regular oscillations as the
myosin activity or contractility of the cell is increased gradually.

{\bf Myosin activation pathways:}
We also performed experiments aimed at understanding the myosin activation
mechanisms involved in the oscillation process. Experiments using drugs to block
the myosin light chain kinase (MLCK)  or the Rho kinase pathway reveals
the following. 
(i) When 25 $\mu$M Y-27632, a Rho kinase inhibitor ~\cite{cohen2000}, is added
to oscillating cells, all the observed cells ($> 100$) stopped oscillating well
within a minute, the time taken for the drug concentration to homogenize.
(ii) Most cells (more than 90\%) suspended in medium without any serum do not
show any shape dynamics. When  50 $\mu$M lisophosphatidic acid (LPA) is
added more than 90\% of the cells become highly dynamic with many showing
oscillatory behavior. LPA activates contractility via the Rho pathway
~\cite{sontheimer1998, sontheimer2000}.
(iii) When exposed to 25 $\mu$M ML-7 to block MLCK ~\cite{hidaka1987}, 
in presence of 1\% serum, all the oscillating cells ($> 100$) stopped
oscillating in about 5 to 10 minutes. The time taken appears
to depend on the amount of serum. When 25 $\mu$M ML-7 was added to cells in
medium without serum and with 50 $\mu$M LPA, the action of the drug was
dramatically faster. Almost all cells ($>$ 95\%)  stopped oscillating within the
first minute and a few that continued stopped in 5 minutes.

The above experiments clearly show that both the Rho kinase as well as the 
MLCK pathways are necessary for the oscillatory dynamics. Further, it shows
that oscillations can be triggered in these cells purely by increasing the Rho
kinase activity using LPA.

\begin{figure}[h]
\includegraphics[width=9.8cm]{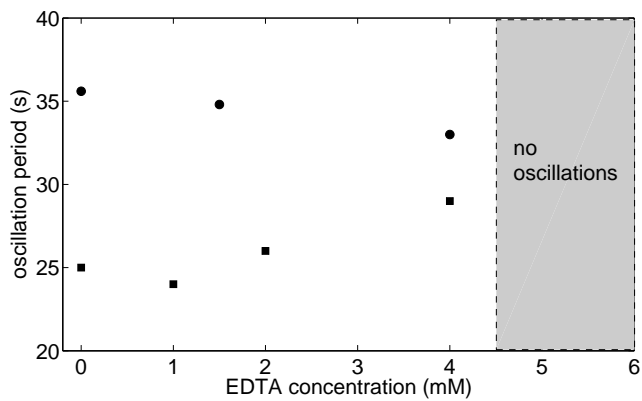}
\vspace{-0.3cm}
\caption{\label{fig:freq_calcium}
Extracellular calcium concentration does not affect the period
in a significant way (different symbols are for different cells).
Above 4 mM EDTA, the oscillations ceased. Oscillations could be
restarted in the same cells by addition of CaCl$_2$ to the medium.
}
\vspace{-0.0cm}
\end{figure}

{\bf Effect of extracellular calcium on oscillations:}
It is known that transient variations in free calcium ion concentration inside
the cell can lead to a modulation of the mechanical properties of the cortical
actin gel ~\cite{alberts}. Such Ca$^{++}$ transients could be triggered as a
result of small influxes of extracellular calcium. In our experiments, when the
external Ca$^{++}$ concentration was gradually depleted from 
its normal concentration of about 1--2 mM using EDTA there was no substantial
change in frequency as shown in Fig.\ \ref{fig:freq_calcium}. At an EDTA
concentration of 4 mM or above, sufficient to
remove all of the extracellular Ca$^{++}$, the oscillations ceased for all the
observed cells. When Ca$^{++}$ was subsequently replenished using
CaCl$_2$, the oscillations resumed, clearly showing that the oscillations can be
switched off and on by altering extracellular free Ca$^{++}$ concentration
in a narrow range.
%

In order to explore the mode of action of extracellular Ca$^{++}$ we 
exposed oscillating cells to Gadolinium ions, a
potent blocker of mechanically activated Ca$^{++}$ channels on the 
plasma membrane ~\cite{sachs1992, sachs2000}. Upon exposure to this agent the
oscillations stopped abruptly in all cells, even in presence of normal
concentration of extracellular calcium. This experiment, along with EDTA
experiments, shows that Ca$^{++}$ influx via mechanosensitive ion channels is
vital to the oscillatory mechanism. 

\begin{figure}[h]
\includegraphics[width=10.0cm]{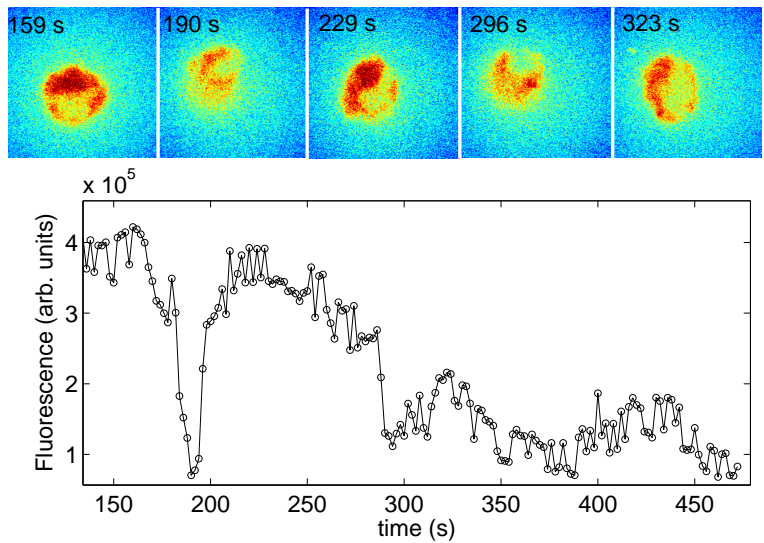}
\vspace{-0.3cm}
\caption{\label{fig:fluorescence}
Images showing the variations in free calcium ion concentration inside a cell
exhibiting strong shape dynamics (the less bright regions inside the cell is
the nucleus), and a plot of the average intensity inside this
cell (also see supplementary files Movie-4 \cite{movies}).
Note that although the hot stage was maintained at 35 ${^\circ}$C, the actual
temperature at the cell was a few degrees lower in this case due to the use of
a high numerical aperture, oil immersion objective, and hence the longer
period of oscillations (about 100 s). After switching on the illumination
photobleaching was too strong from 0--150 s making analysis difficult in this
time segment (see text for details).
}
\vspace{-0.0cm}
\end{figure}
{\bf Intracellular calcium variations:}
When the cells are loaded with Fluo-4 AM dye and observed under a fluorescence
microscope continuous variations in the fluorescence intensity can be observed
in dynamic cells. In some cases the variations appears to be roughly periodic.
Long term observation is not possible due to photobleaching of the dye and the 
resulting toxicity affecting the dynamics. A typical example is shown in 
Fig.\ \ref{fig:fluorescence}. Note that the longer period of oscillation
(about 100 s) in this case is due to the lower temperature at the cell
resulting from the use of an oil immersion objective. Although this suggests an
oscillatory variation of free calcium in the oscillating cells, this result
should be treated a preliminary for the following reasons.
(i) The cytoplasm inside the oscillating cells flows in correlation
with the shape changes. This flow carries with it fluorescently labeled regions
(like calcium rich areas or calcium stores inside the cell). This problem is
greatly reduced by reducing the quantity of dye used as well as the dye loading
time.
(ii) The cells are very sensitive to the excitation frequency of about 488 nm
once they are loaded with the dye. Under strong illumination the cells stop
oscillating or blebbing within a few seconds ($<5$ s). Under weak illumination,
cells can be observed for a few minutes.
(iii) Even though Fluo-4 is widely accepted as an efficient indicator of free
calcium in cells, a recent study finds that the dye, when exposed to the
excitation light, can in itself cause calcium transients ~\cite{bader2003}.

{\bf Bleb dynamics in relation to oscillations:}
%
\begin{figure}[h]
\includegraphics[width=15.0cm]{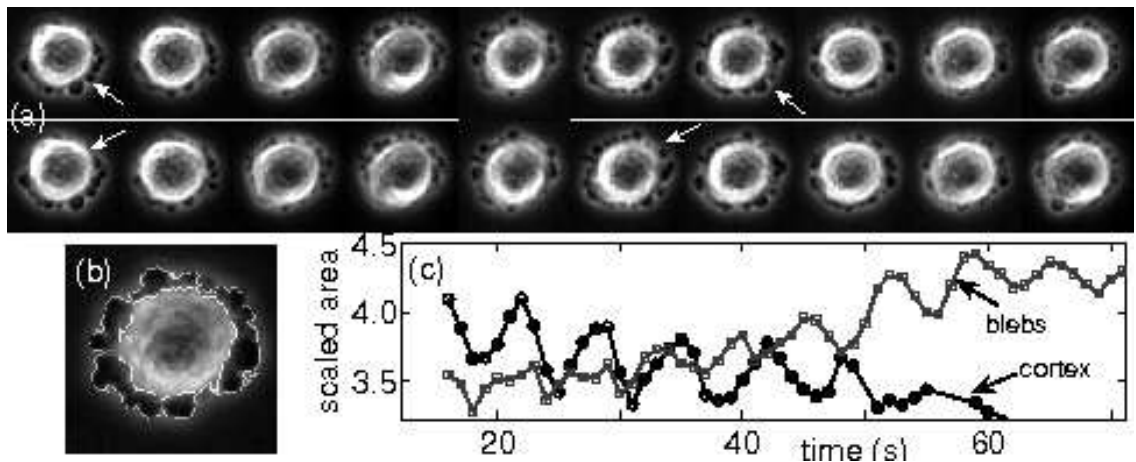}
\vspace{-0.3cm}
\caption{\label{fig:bleb_panel}
(a) Sequence of images showing blebbing dynamics in a cell which also shows
shape oscillations. In this particular case an untreated glass substrate was
employed, and the cell eventually attaches and spreads on the glass.
Initially, blebbing appears to be roughly periodic with maxima at frames
indicated by the arrows. (b) A rough measure of blebbing is obtained by
detecting the envelope which includes all the blebs (outer boundary). The
cortical region can also be detected due to the higher intensity of this region
in phase contrast images (inner boundary). (c) Oscillations
of the area of the envelope of blebs (squares) and area of cortex (circles). The
centroid of the detected cortex also oscillates. The plots show that shape
oscillations and blebbing can coexist and blebbing can be roughly periodic in
some cases.
}
\vspace{0.6cm}
\end{figure}
It is observed that about 50\% of the oscillating cells exhibit moderate
blebbing or membrane bulge formation while showing shape oscillations. 
About a few percent can exhibit severe blebbing or blebbing without
oscillations (see supplementary movie Movie-5 \cite{movies}). Often a cell
oscillates for a
while without any signs of membrane blebs or bulge formation and then begins to
exhibit small localized bleb formation which correlates with the oscillations
(see supplementary movie Movie-6 \cite{movies}). Detailed experiments on cell
oscillations and
blebbing demonstrate a remarkable set of similarities between these two
processes. 
(i) They are both dynamical processes requiring acto-myosin contractility
as shown by treatments using blebbistatin to inhibit myosin or with
Latrunculin-A to disrupt actin filaments. In suspension, both these shape
dynamics can occur continuously for several hours without any significant change
in their rates. 
(ii) Some cells show both periodic oscillations and blebbing simultaneously.
In rare cases, the extent of blebbing appears to be roughly periodic with
a periodicity comparable to that of the oscillations as shown in
Fig.\ \ref{fig:bleb_panel} (also see supplementary movie Movie-7 \cite{movies}).
(iii) More remarkably, a cell can spontaneously get into one of these two 
modes or even switch from one to the other as a function of time.
(iv) Removal of extracellular calcium using  EDTA or exposure to the
Ca$^{++}$ channel blocker gadoliniun ions immediately stops blebbing as well as
oscillations. Introducing Ca$^{++}$ after EDTA treatment reinitiates blebbing
or oscillations in the same cells.
(v) All the drug and LPA experiments mentioned earlier give the same
results for both the blebbing as well as the oscillatory dynamics.

The oscillatory shape dynamics and the blebbing dynamics are separated by time
and length scales. The typical time scale for bleb growth and disappearance,
which is about 10 s, is shorter compared to 40 s for oscillations at
the same temperature. Similarly, blebbing occurs with a typical separation
between blebs of a few microns,  about 5--10 times shorter than the cell size.
Blebbing dynamics shows a wide distribution of these time and length scales
while the oscillations are periodic and global. Further, membrane bulge
formation, which characterizes blebbing, occurs only in a subset of oscillating 
cells. However, despite these differences, the above listed
experiments suggest that the underlying mechanisms responsible for cell
oscillations and cell blebbing dynamics are essentially the same. More
specifically, both dynamical behaviors arise from the inherent contractile
nature of the actomyosin gel and involve Ca$^{++}$ signaling via stretch
activation of calcium channels. This point will be further elaborated in the 
following section. 

\section*{DISCUSSION}

Oscillatory shape dynamics can be induced in a variety of otherwise
non-oscillatory cells, usually when they are exposed to toxins that
specifically depolymerize microtubules. Bornens et al. ~\cite{bornens1989}
observed that lymphoblasts with disrupted microtubules exhibit an actomyosin 
constriction ring which propagates back and forth with a
constant velocity of about 0.1 $\mu{\rm m}s^{-1}$. Subsequently,
Pletjushkina et al. ~\cite{pletjushkina2001} reported transient periodic
cortical oscillations occurring in spreading fibroblasts when microtubules
are disrupted. Treated cells also exhibited strong blebbing. Further, they
demonstrated a correlation between shape oscillations and free Ca$^{++}$
concentration inside the cell. Calcium influx could also be imaged in large
blebs. More recently, Paluch et al. ~\cite{paluch2005} investigated oscillatory 
dynamics in cell fragments occurring when microtubules are disrupted. It is
shown that the actin cortex in fragments can break as a result of
excess tension. This leads to an oscillatory dynamics which involves the
repeated formation of a bare membrane bulge and a contractile ring which
propagates back and forth in correlation with the oscillations. 

Unlike the above examples, the oscillatory dynamics reported here do not
involve any biochemical treatments. Non-adherent cells are observed to
oscillate spontaneously. Since myosin-II and actin are necessary for this
dynamics, it points to an inherent instability of the actin cortex in these
cells when the contractile cortex is unconstrained by cell-substrate adhesion.
Moreover, the oscillations are very periodic and can be sustained for many
hours. The experiments show that myosin phosphorylation via the calcium
dependent MLCK pathway as well as the calcium independent Rho kinase
pathway are involved and necessary. The oscillatory dynamics is also
coupled to extracellular calcium, which enters the cell via stretch activated
calcium channels. The oscillation frequency is independent of the external
calcium concentration except when the external calcium is almost completely
depleted. But there is a clear increase(decrease) in frequency when myosin
activity or cell contractility is increased(decreased) gradually. While
some cells show moderate to severe blebbing, these cells can also oscillate
without any observable bleb formation or membrane bulge formation. This
suggests that in such cells the cortex and the membrane-cortex adhesion remain
mostly intact during oscillations.

\begin{figure}[h]
\includegraphics[width=10.0cm]{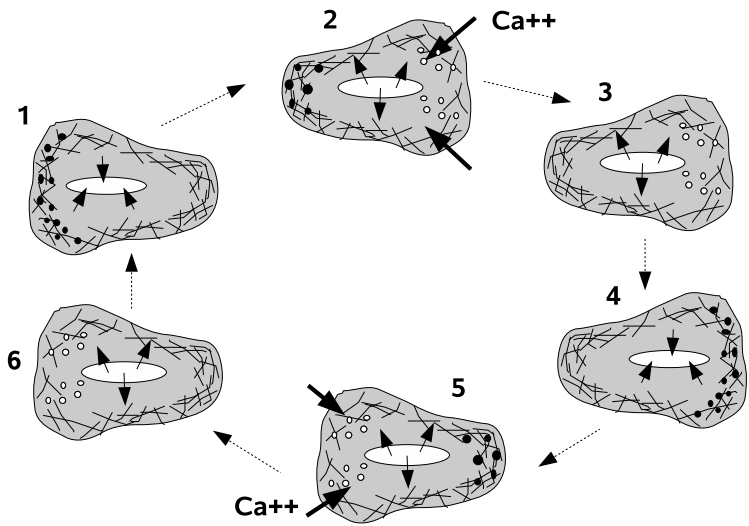}
\vspace{-0.3cm}
\caption{\label{fig:schematic}
A schematic of the dynamics explained in the text. Calcium increase is
indicated by tiny white circles and phosphorelated myosin by black patches. The
actin cortex contracts anisotropically (1); the membrane and cortex at the
opposite end is stretched and causes calcium influx through stretch activated
channels (2); if the Ca$^{++}$ level raises high enough, the gel may partially
solate due to the excess calcium (3); as calcium is reabsorbed into internal
stores, actin polymerizes, myosin phosphorelates and binds, and the gel
contracts (4); the cycle repeats
}
\vspace{-0.0cm}
\end{figure}

Based on these observations, we propose a possible mechanisms for oscillations
of the actomyosin cortex which involve a coupling to stretch activated calcium
channels on the outer membrane. Due to the presence of myosin motors, the
cortical gel is contractile and generates lateral tension. In a general case,
this tension is anisotropic either due to a non-uniform distribution of myosin
or due to anisotropies in the gel structure. This leads to an anisotropic shape
change of the cortex. If the shape change is significant, it can lead to a
stretching of the membrane due to a change in the hydrostatic pressure inside
the cell. The cytoplasmic streaming which is observed suggests such gradients in
pressure (Movie-8 \cite{movies}). The tension thus generated opens tension
sensitive calcium channels
causing an influx of calcium ions into the cell. Tension on actin filaments
may also play a role in activating gadolinium-sensitive calcium channels as
has been reported for fibroblasts ~\cite{mcculloch1994}. The free calcium ion
concentration inside a cell is about $10^{-4}$ mM is far lower than that of its
extracellular value of 1--2 mM ~\cite{alberts}. The entry of calcium from
outside can thus lead to a significant increase in the free calcium ion
concentration inside the cell. This calcium influx may also trigger a
calcium-induced calcium release from internal calcium reserves. Such a process 
has been observed in fibtroblasts and other non-muscle cells in response to
mechanical stretch ~\cite{mcculloch1994, jacobson1999}, although this may
depend on cell type ~\cite{mcculloch2001, huang2005}. Subsequently, the calcium
concentration is brought back to the rest value by absorption of free calcium
and by the action of calcium pumps ~\cite{alberts}. The initial rapid elevation
of free calcium has mainly
two effects on the cortical actomyosin gel. Large increases in free calcium
concentration can lead to a partial solation or weakening of the gel due to
gelsolin mediated severing of actin filaments and can alter the actin
polymerization rates ~\cite{alberts, bray, taylor1991, stossel1993}. Calcium
also affects actin crosslinkers like $\alpha$-actinin and other actin
binding proteins leading to a weakening of the gel at higher concentrations
($\sim 10^{-3}$ mM, ref. ~\cite{fechheimer2002}, and references
therein). The extent of solation depends on the calcium concentration.
In-vitro studies show that a low degree of solation can increase the
contraction rates and strong solation can make the gel non-contractile
~\cite{taylor1991}. At low concentrations, calcium also causes myosin
phosphorylation via the calmodulin-MLCK pathway, by which myosin is assembled
into contractile filaments ~\cite{alberts, bray, suzaki2006, lee2004,
jacobson1999}. This enhances the contractility of the cortex. These two
processes--the weakening of the gel and increased contractility--may be
separated in time. At high calcium levels the gel is partially solated and
as the calcium is reabsorbed into the stores, actin polymerizes, myosin
filaments bind and the cortex become contractile again. The whole cycle is then
repeated. This scenario, where repeated contractions occur with calcium playing
a feedback role is schematically represented in Fig.\ \ref{fig:schematic}. 
%

It is not clear from the present experiments as to what extent the actin gel
structure is altered by the increase in calcium levels. If severe calcium
mediated disruption of the cortex occurs, it will have the effect of increasing
the time delay between calcium influx and the initiation of contractile
response, as the cortex has to reform before contraction can take effect. The
role of calcium in effecting myosin phosphorylation is more clear as shown by
experiments using drugs that block the MLCK pathway. However, experiments show
that the frequency of oscillations do not depend on the extracellular calcium
levels even when external calcium levels are reduced using EDTA. This may be
understood if a small influx of extracellular calcium is enough to trigger a
much larger calcium-induced calcium release from internal stores. 

It is observed that cells which have stopped oscillating after external calcium
is completely chelated using EDTA can restart oscillations if calcium
is reintroduced in the form of CaCl$_2$. How this calcium enters the cell and
triggers oscillations in a non-oscillating cell is not clear. However, it may be
noted that calcium channels on the membrane can open stochastically and generate
``calcium puffs'' ~\cite{dawson2006}. Such stochastic Ca$^{++}$ puffs may aid in
``kick starting'' oscillations by inducing contraction when Ca$^{++}$ is added
from outside.


Experiments clearly demonstrates that the frequency of oscillation increases
with increasing myosin activity. The maximum observed variation is about 30\%
in the blebbistatin experiment. This may be understood as a direct effect of
the rate of contraction of the gel, which is expected to increase with
increasing number of active myosin motors as observed in in-vitro gels
~\cite{taylor1991}. The fact that the oscillation ceases at a finite frequency
in the blebbistatin and serum experiments indicates the existence of a threshold
in myosin activity below which an oscillatory instability cannot
occur. Moreover, the observation that myosin phosphorylation via both Rho kinase
as well as the
MLCK pathways are necessary further suggests that a high level of phosphorelated
myosin is necessary for this dynamics. Indeed, crosslinked in-vitro gels too
exhibit such a threshold as a function of myosin-II concentration
~\cite{taylor1991}. In earlier examples of oscillations induced by microtubule
depolymerization in spreading fibroblasts, the Rho kinase activation was
proposed as the triggering mechanism for oscillations ~\cite{pletjushkina2001}.
This enhanced contractility may have been necessary as these cells are attached
and spread on a substrate and hence the cell cortex is partly constrained and
under tension. In our case, we observe that the oscillations become
progressively weaker in amplitude and die out completely as cell spreading
proceeds, supporting this hypothesis. Thus, detachment of cells from
the substrate may have the effect of lowering this contractility threshold.

In Paluch et al. ~\cite{paluch2005}, observations made on cell fragments clearly
show breakage of the actin cortex indicating large tension. The cortical
breakage is accompanied by a membrane bulge formation. This may be because the
bare membrane is unable to sustain the excess hydrostatic pressure generated due
to contraction. In the present case, cells showing periodic oscillations often
show no membrane bulge formation or blebbing. This is a strong indication that
the actin cortex as well as the membrane-cortex adhesion are intact during
oscillations. Indeed, in some cases, a cell can oscillate without any membrane
bulge formation for tens of minutes (observation time) and then suddenly
developed repeated bulge formation in correlation with oscillations. An example
is shown in Movie 5. In such cases either the cortex has become locally weak and
tend to break during contraction or the membrane-cytoskeletal linkages become
locally weak resulting in membrane detachment and bleb formation.

Unlike shape oscillations, blebbing dynamics has been investigated in detail
(see, for example, C. Cunningham ~\cite{cunningham1995-bleb}) and mechanisms
have been proposed recently ~\cite{charras2005, dobereiner2006}. It is shown
that repeated local contractions of the actomyosin complex occurs during bleb
formation. According to Charras et al. ~\cite{charras2005}, the contractions
generate a pressure which is diffusive due to the resistance to flow between the
cytoplasm and the cortical gel. This produces local maxima in pressure which
causes the membrane to detach and form blebs, at a constant cell volume. As
mentioned earlier, in rare cases (a few percent), cells can be observed to
switch between seemingly random blebbing dynamics to regular oscillatory
dynamics as a function of time in the range of tens of minutes. Moreover,
experiments show that actomyosin contractility and similar calcium signaling are
involved in both these dynamical processes.

Why do some cells show periodic and global shape oscillations (Movie-1) while
some others show more complex contractile behavior (Movie-3) and some show
blebbing (Movie-5)? A possible answer to this question may lie in the
viscoelastic and contractile nature of the actomyosin gel. Stresses
generated locally
in such a gel will typically relax over a certain length and time scale which
depends on the extent of crosslinking and hence the elasticity of the gel.  If
the gel is highly crosslinked (more elastic), for example, stresses may
propagate distances comparable to the cell size. Therefore, any contractile
stresses are felt all over the cell. This could lead to correlated shape
alterations as is observed in periodically oscillating cells. If on the other
hand, the gel is elastically weak (more viscous than elastic) the stresses die
out very quickly, over distances considerably smaller than the cell size. Volume
elements of the gel which are separated by distances which are bigger than this
length scale will not feel each other's mechanical state. Thus the cell could
sustain several local contractions or relaxations which are largely
uncorrelated. Such a transformation from global contraction to local
small-scale contractions is observed in in-vitro gels as the extent of
crosslinking is reduced ~\cite{taylor1991}. Local contractions could then lead
to blebbing due to local build up of pressure and generate local membrane
tension ~\cite{charras2005}. The viscoelastic properties of the gel may vary
from cell to cell or during a cell cycle giving rise to a range of dynamics in a
given cell population. The gel properties may also be modulated over time either
due to effects of calcium or due to other complex biochemical events and the
cell can switch from one mode of dynamics to another. The timescale for bleb
formation and retraction will, in general, depend on the bleb size and the time
taken for a fresh cortex to form below the bare membrane and induce contraction
~\cite{charras2005}.
In fact, this is akin to the shape dynamics in cell fragments where cortical
breakage and membrane bulge formation occurs ~\cite{paluch2005}. In
periodically oscillating cells no such membrane bulge formation is evident and
this may explain why the oscillations become remarkably regular.


Finally, it is interesting to note that periodic dynamics and propagating waves
are observed in the lamellipodea of a wide variety of spreading or
locomoting cells
~\cite{xenias-sheetz2004, danuser2006, sheetz2004, sheetz2006}, 
and in cells recovering from actin depolymerization ~\cite{anderson2004}.
The wave-like cortex dynamics observed in some of the oscillating 
cells (see supplementary movie Movie-2) is reminiscent of the lamellipodial
waves observed in some of these cases ~\cite{danuser2006, sheetz2006}.

{\bf Conclusion:}
We show that fibroblasts can exhibit sustained, remarkably periodic, shape
oscillations when maintained in suspension, without any treatment
with drugs. Even cells removed from the culture by gentle shaking show
sustained oscillatory behavior. These cells are, in general, close to
the mitotic state and hence loosely adherent. Thus, the observed dynamics
is fundamental to fibroblast mechanics and may well be relevant under
physiological conditions. The implications of this phenomenon to cytokinesis
~\cite{spudich2004}, cell locomotion ~\cite{kobatake1985, ewa_rev}, or during
embryogenesis when the cells are loosely adherent will be of interest for future
research. The robustness of the oscillations and the fact that it can be
sustained for several hours makes it an exceptional and relatively simple model
system for experiments aimed at understanding the dynamical properties of
cortical actin gel. The results presented here are also expected to motivate
future theoretical modeling based on the mechanics of contractile actomyosin
gels.

{\bf Acknowledgments:}
P. A. P would like to thank Albrecht Ott for his generous support and for a
critical reading of the manuscript, Jacques Prost and Jean-Fran\c{c}ois
Joanny for encouragement and several stimulating discussions, and Guillaume
Salbreux for valuable discussions. Thanks are also due to C\'ecile Sykes for her
constructive criticisms and suggestions, to Ewa Paluch and an anonymous referee
for suggesting the exploration of the myosin activation pathways, to Pablo
Fernandez and Jordi Soriano for useful suggestions, and to Andrea Hanold for
technical support. The funding for this project was provided by the University
of Bayreuth.

\bibliography{pullarkat_bibliography}

\end{document}